\documentclass[12pt]{article}
\usepackage{enumerate}
\usepackage[hyperindex,breaklinks,colorlinks,citecolor=blue,pagebackref]{hyperref}
\usepackage[latin1]{inputenc}
\usepackage{amsmath, amsthm, amsfonts, amssymb}
\usepackage{mathrsfs}
\usepackage{latexsym}
\usepackage{fullpage}
\usepackage[mathcal]{eucal}
\usepackage[all]{xy}
\usepackage{palatino}

\newtheorem{lemma}{Lemma}[section]
\newtheorem{proposition}{Proposition}[section]
\newtheorem{theorem}{Theorem}[section]
\newtheorem{corollary}{Corollary}[section]

\theoremstyle{definition}

\theoremstyle{remark}

\newtheorem{rmks}{Remarks}[section]

\newtheorem*{Examples}{Examples}

\newcommand{\defin}[1]{\emph{\textbf{#1}}}

\newcommand{\N}{\mathbb{N}}

\newcommand{\R}{\mathbb{R}}

\newcommand{\Kucera}{Ku$\mathrm{\check{c}}$era}

\newcommand{\ML}{\mathsf{ML}} 

\newcommand{\cantor}{\{0,1\}^\N}

\newcommand{\integral}[2]{\int\!{#1}\,\mathrm{d}{#2}}

\newcommand{\Dim}{\mathrm{Dim}}

\newcommand{\rst}[1]{\ensuremath{{\mathbin\upharpoonright}\raise-.2ex\hbox{\scriptsize $#1$}}}

\title{The dimension of ergodic random sequences}
\author{Mathieu Hoyrup}

\begin{document}
\maketitle
\begin{abstract}
Let $\mu$ be a computable ergodic shift-invariant measure over $\{0,1\}^\N$. Providing a constructive proof of Shannon-McMillan-Breiman theorem, V'yugin proved that if $x\in\{0,1\}^\N$ is Martin-L\"of random w.r.t. $\mu$ then the strong effective dimension $\Dim(x)$ of $x$ equals the entropy of $\mu$. Whether its effective dimension $\dim(x)$ also equals the entropy was left as an problem question. In this paper we settle this problem, providing a positive answer. A key step in the proof consists in extending recent results on Birkhoff's ergodic theorem for Martin-L\"of random sequences.
{}\\

\end{abstract}

\section{Introduction}

The \emph{effective dimension} and \emph{strong effective dimension} of an infinite binary sequence $x$ are defined as
\begin{align*}
\dim(x) & = \liminf_n\frac{K(x\rst{n})}{n} \\
\Dim(x) & =\limsup_n\frac{K(x\rst{n})}{n},
\end{align*}
where $K(w)$ is the Kolmogorov complexity of $w$.

They can be characterized as effective versions of Hausdorff and packing dimensions respectively, or by divergence of $s$-gales (see \cite{Lutz00, May02, Ath07} for the original results and \cite{Lutz05} for a survey).

Let $p\in[0,1]$ be a computable real number and $\mu_p$ the Bernoulli measure over Cantor space given by $\mu_p[w]=p^{|w|_1}(1-p)^{|w|_0}$. It is well-known that if an infinite binary sequence $x$ is Martin-L\"of random w.r.t. $\mu_p$ then $\dim(x)=\Dim(x)=h(\mu_p)$, where $h(\mu_p)$ is the entropy of $\mu_p$ defined by
\begin{equation}\label{entropy_bernoulli}
h(\mu_p)=-p\log(p)-(1-p)\log(1-p).
\end{equation}

This result is not difficult to prove and reduces to the strong law of large numbers for Martin-L\"of random sequences, as on the one hand\footnote{$K$ is the prefix version of Kolmogorov complexity}
\begin{equation*}
K(x\rst{n})=-\log \mu_p[x\rst n]+O(1)
\end{equation*}
for $\mu_p$-random sequences by Levin-Schnorr theorem, and on the other hand
\begin{equation*}
-\frac{1}{n}\log \mu_p[x\rst{n}]=-\frac{|x\rst{n}|_1}{n}\log(p)-\frac{|x\rst{n}|_0}{n}\log(1-p)
\end{equation*}
which converge to $h(\mu_p)$ for $\mu_p$-random sequences, by the Strong Law of Large Numbers for Martin-L\"of random sequences.

This result highlights the relationship between Shannon's information theory, Kolmogorov algorithmic information theory and effective randomness.

Ergodic theory provides a natural extension of information theory in which many results can be transferred, with more involved proofs, from the case of independent identically distributed random variables to the ergodic case, where independence is only required \emph{asymptotically, in the average} (see Section \ref{sec_back} for a precise definition).

First, the strong law of large numbers extends to Birkhoff's ergodic theorem. Second, the coincidence between local information and entropy extends through the Shannon-McMillan-Breiman theorem. Whether Martin-L\"of randomness fits with these theorems has been an open problem for a while. The first results were proved by V'yugin \cite{Vyu98}, based on non-classical, constructive proofs of the theorems. He proved, in particular:

\begin{theorem}[Effective Birkhoff ergodic theorem I]\label{thm_birkhoff_vyugin}
Let $\mu$ be a computable shift-invariant ergodic measure over $\{0,1\}^\N$ and $f\in L^1(\mu)$ be computable. For every Martin-L\"of $\mu$-random sequence $x$,
\begin{equation*}
\lim_{n\to\infty}\frac{1}{n}\sum_{k=0}^{n-1}f\circ T^k(x)=\integral{f}{\mu}.
\end{equation*}
\end{theorem}

The entropy of an ergodic measure is defined as
\begin{equation}\label{eq_entropy2}
h(\mu) = \lim_{n\to\infty}-\frac{1}{n}\sum_{|w|=n}\mu[w]\log \mu[w].
\end{equation}
Observe that (\ref{entropy_bernoulli}) and (\ref{eq_entropy2}) are consistent as they give the same quantity when $\mu$ is a Bernoulli measure.

\begin{theorem}[Effective Shannon-McMillan-Breiman theorem I]\label{thm_sh1}
Let $\mu$ be a computable shift-invariant ergodic measure over $\{0,1\}^\N$. For every Martin-L\"of $\mu$-random sequence $x$,
\begin{equation*}
\limsup_{n\to\infty}\frac{K(x\rst{n})}{n}=\limsup_{n\to\infty}-\frac{1}{n}\log \mu[x\rst{n}]=h(\mu).
\end{equation*}
\end{theorem}
The question whether $\liminf\frac{K\rst{n}}{n}$ coincides with $h(\mu)$ for every Martin-L\"of $\mu$-random was left open by V'yugin. An alternative proof of Theorem \ref{thm_sh1} approximating ergodic measures by Markovian measures was later developed by Nakamura \cite{Nak05}, but also left the question open. In this paper we provide a positive answer to this question.


A classical proof of the Shannon-McMillan-Breiman theorem uses Birkhoff's ergodic theorem, applied to some particular functions. The problem in making it effective is that these functions are not computable in general. Recent works have been achieved to push the effective ergodic theorem to the largest possible class of functions. Here we extend it enough to get the full effective Shannon-McMillan-Breiman theorem.

In Section \ref{sec_back} we recall basic notions of computability, randomness and ergodic theory. In Section \ref{sec_erg} we develop effective versions of Birkhoff's ergodic theorem. In Section \ref{sec_sh} we present our main result.

\section{Background and notations}\label{sec_back}
We work on the Cantor space $\cantor$ of infinite binary sequences. A finite word $w\in\{0,1\}^*$ determines the cylinder $[w]\subseteq\cantor$ of infinite sequences starting with $w$. If $x\in\cantor$ and $n\in\N$, $x\rst{n}$ is the prefix of $x$ of length $n$, and is also denoted $x_0x_1\ldots x_{n-1}$. The cylinders form a base of the product topology.

\paragraph{Effective topology.} An open set $U\subseteq\cantor$ is \defin{effective} if it is a recursively enumerable union of cylinders. A closed set is effective it its complement is an effective open set. A function $f:\cantor\to\R$ is \defin{computable} if there is a Turing machine that on oracle $x$ and input $n$ computes a rational number $q$ such that $|q-f(x)|<2^{-n}$. Equivalently, $f$ is computable if for every rational numbers $a<b$, $f^{-1}(a,b)$ is effectively open, uniformly in $a,b$. A function $f:\cantor\to[0,+\infty]$ is \defin{lower} (resp. \defin{upper}) \defin{semi-computable} if there is a Turing machine that on oracle $x$ and input $n$ computes a rational number $q_n$ such that $f(x)=\sup_n q_n$ (resp. $f(x)=\inf_n q_n$). Equivalently, $f$ is lower (resp. upper) semi-computable if for every rational number $a$, $f^{-1}(a,+\infty]$ (resp. $f^{-1}[0,a)$) is effectively open, uniformly in $a$.

\paragraph{Kolmogorov complexity and Martin-L\"of randomness.} For $w\in\{0,1\}^*$, $K(w)$ is the prefix version of Kolmogorov complexity, defined by Levin and Chaitin independently. It is defined as the length of a shortest input of a universal Turing machine with prefix-free domain computing $w$ on that input.

A probability measure $\mu$ over $\cantor$ is determined by its value on cylinders $\mu[w]$, for $w\in\{0,1\}^*$. $\mu$ is computable if $\mu[w]$ is a computable real number, uniformly in $w$. Given a computable probability measure $\mu$, a sequence $x\in\cantor$ is \defin{Martin-L\"of $\boldsymbol\mu$-random}, denoted $x\in\ML_\mu$, if there is $c$ such that for all $n$,
\begin{equation*}
K(x\rst{n})\geq -\log \mu[x\rst{n}]-c.
\end{equation*}
Martin-L\"of's original definition \cite{MLof66} was expressed in terms of tests rather than complexity, but the one given here, due to Levin and Chaitin \cite{Cha75} independentely, was proved to be equivalent by Levin \cite{Lev73} and Schnorr \cite{Sch73}.

The function
\begin{equation*}
t_\mu(x)=\sup_n\{-\log \mu[x\rst{n}]-K(x\rst{n})\}
\end{equation*}
is lower semi-computable and $\integral{2^{t_\mu}}{\mu}\leq 1$. Moreover, it was proved in \cite{Gac80} that $2^{t_\mu}$ is maximal in the sense that for every integrable lower semi-computable function $f:\cantor\to[0,+\infty]$, there exists $c_f$ such that $f\leq c_f2^t_\mu$. We call such an $f$ a \defin{$\boldsymbol\mu$-test}. It tests Martin-L\"of randomness in the sense that $x\in\ML_\mu$ iff $f(x)<\infty$ for each $\mu$-test $f$ iff $t_\mu(x)<\infty$. More can be found on this subject in \cite{Nies09, dowhir10}.

\paragraph{Ergodic theory.}
We recall some basic notions of ergodic theory, more details can be found in \cite{Smo71, Pet83}. We denote by $T:\cantor\to\cantor$ the \defin{shift map} defined by $T(x_0x_1\ldots)=x_1x_2\ldots$. A measure $\mu$ over $\cantor$ is \defin{shift-invariant} if for all Borel sets $A$, $\mu(T^{-1}A)=\mu(A)$, equivalently if $\mu[0w]+\mu[1w]=\mu[w]$ for all $w\in\{0,1\}^*$. $\mu$ is \defin{ergodic} if for all Borel sets $A$ such that $T^{-1}A=A$ up to a null sets, $\mu(A)=0$ or $1$. Equivalently, $\mu$ is ergodic if for all $u,v\in\{0,1\}^*$,
\begin{equation*}
\lim_{n\to\infty}\frac{1}{n}\sum_{k=0}^{n-1}\mu([u]\cap T^{-k}[v])=\mu[u]\cdot \mu[v].
\end{equation*}

\section{Effective ergodic theorems}\label{sec_erg}
The following theorem, taken from \cite{BienvenuDMS10}, extends a result of \Kucera~from the uniform measure to any ergodic shift-invariant measure:

\begin{theorem}[Effective Poincar\'e recurrence theorem]\label{thm_poincare}
Let $\mu$ be a computable ergodic shift-invariant measure and $C\subseteq \cantor$ an effective closed set such that $\mu(C)>0$. Every Martin-L\"of $\mu$-random sequence has a tail in $C$, i.e. for every $x\in\ML_\mu$ there exists $k$ such that $T^k(x)\in C$.
\end{theorem}

In \cite{BieHoyShen10} and  \cite{FraGreeMillNg10} independently this result was used to prove that not only the orbit of $x$ eventually falls into $C$, but it does so with frequency $\mu(C)$.

\begin{theorem}[Effective Birkhoff ergodic theorem II]\label{thm_brk_lsc}
Let $\mu$ be a computable ergodic shift-invariant measure and $C\subseteq \cantor$ an effective closed set such that $\mu(C)>0$. For every Martin-L\"of $\mu$-random sequence $x$,
\begin{equation*}
\lim_{n\to\infty}\frac{1}{n}|\{k<n:T^k(x)\in C\}|=\mu(C).
\end{equation*}
\end{theorem}

We first generalize the result from sets to functions:
\begin{theorem}[Effective Birkhoff ergodic theorem III]\label{thm_erg_semi}
Let $\mu$ be a computable ergodic shift-invariant measure. Assume $f:\cantor\to [0,+\infty]$ is:
\begin{itemize}
\item either lower semi-computable,
\item or upper semi-computable and bounded by a $\mu$-test.
\end{itemize}
For each $x\in\ML_\mu$,
\begin{equation*}
\lim_{n\to\infty}\sum_{k=0}^{n-1}f\circ T^k(x)=\integral{f}{\mu}.
\end{equation*}
\end{theorem}
\begin{proof}
Let us introduce the notation $A_n^f(x)=\frac{1}{n}(f(x)+\ldots+f\circ T^{n-1}(x))$.

If $f$ is lower semi-computable, then there is a sequence of uniformly computable nonnegative functions $f_n\nearrow f$. Applying Theorem \ref{thm_birkhoff_vyugin} to $f_n$ and $x\in\ML_\mu$ gives $\liminf_k A_k^f(x)\geq \liminf_k A_k^{f_n}(x)= \integral{f_n}{\mu}$. By the monotone convergence theorem, $\integral{f_n}{\mu}\nearrow\integral{f}{\mu}$, so $\liminf_k A_k^f(x)\geq \integral{f}{\mu}$. If $\integral{f}{\mu}=\infty$ we are done. Otherwise, let $q>\integral{f}{\mu}$ be a rational number. The set $C_K:=\{x:\forall k\geq K, A_k^f(x)\leq q\}$ is effectively closed and by the classical ergodic theorem, there exists $K$ such that $\mu(C_K)>0$. Theorem \ref{thm_poincare} tells us that if $x\in\ML_\mu$ then there is $n$ such that $T^n(x)\in C_K$. As a result, $\limsup A_k^f(x)=\limsup A_k^f(T^n(x))\leq q$. As this is true of every $q>\integral{f}{\mu}$, we get the result.

Now, if $f$ is upper semi-computable and $f\leq t$ where $t$ is a $\mu$-test, then for $x\in\ML_\mu$, applying the preceding result to $t$ and $t-f$,
\begin{equation*}
A^f_n(x) = A^t_n(x)-(A^{t-f}_n(x))  \to \integral{t}{\mu}-\integral{(t-f)}{\mu} = \integral{f}{\mu}.
\end{equation*}
\end{proof}

We then extend this result further:

\begin{corollary}[Effective Birkhoff ergodic theorem IV]\label{cor_delta2_bounded}
Let $f:\cantor\to[0,+\infty]$ be $\Delta^0_2$ on $\ML_\mu$, i.e. there is a sequence $f_n$ of uniformly computable functions such that $f(x)=\lim_n f_n(x)$ for each $x\in\ML_\mu$. Assume that $f$ is dominated by a $\mu$-test. For every $x\in\ML_\mu$,
\begin{equation*}
\lim_{n\to\infty}\sum_{k=0}^{n-1}f\circ T^k(x) =\integral{f}{\mu}.
\end{equation*}
\end{corollary}

\begin{proof}
Let $g_N=\inf_{n\geq N}f_n$ and $h_N=\min(t,\sup_{n\geq N}f_n)$. On $\ML_\mu$, $g_N\nearrow f$ and $h_N\searrow f$. By the monotone and dominated convergence theorem, the convergences hold in $L^1(\mu)$. Applying Theorem \ref{thm_brk_lsc} to $g_N$ and $h_N$ gives the result. More precisely, for every $x\in\ML_\mu$ and every $N$,
\begin{align*}
\liminf_n A_n^f(x) & \geq \liminf_n A_n^{g_N}(x)=\integral{g_N}{\mu} \\
\limsup_n A_n^f(x) & \leq  \limsup_n A_n^{h_N}(x)=\integral{h_N}{\mu},
\end{align*}
so
\begin{equation*}
\integral{f}{\mu}=\sup_N\integral{g_N}{\mu}\leq\liminf_n A_n^f(x)\leq \limsup_n A_n^f(x)\leq\inf_N\integral{h_N}{\mu}=\integral{f}{\mu}.
\end{equation*}
\end{proof}

\section{The effective Shannon-McMillan-Breiman theorem}\label{sec_sh}

We now present our main result.

\begin{theorem}[Effective Shannon-McMillan-Breiman theorem II]\label{thm_main}
Let $\mu$ be a computable shift-invariant probability measure. For each $x\in\ML_\mu$,
\begin{equation*}
\lim_{n\to\infty}\frac{K(x\rst{n})}{n} = \lim_{n\to\infty}-\frac{1}{n}\log \mu[x\rst{n}] = h(\mu).
\end{equation*}
\end{theorem}

A proof of the classical result, stating the result for a.e. $x$, can be found in \cite{Smo71, Pet83}. It makes use of martingale convergence theorems and ergodic theorems. The main difficulty in adapting the proof is to make sure that the effective versions of the ergodic theorem can be applied. The rest of this section is devoted to the proof of Theorem \ref{thm_main}. 

An easy calculation shows that
\begin{equation}\label{eq0}
-\log \mu[x\rst{n}] = \sum_{k=0}^{n-1}f_{n-1-k}\circ T^{k}(x)
\end{equation}
where
\begin{alignat*}{2}
f_k(x) & :=  -\log \mu[x_0|x_1\ldots x_k]=-\log\frac{\mu[x_0\ldots x_k]}{\mu[x_1\ldots x_k]}&\quad& \text{for }k\geq 1,\\
f_0(x) & :=  -\log \mu[x_0].
\end{alignat*}

\begin{lemma}
$f_k(x)$ converge for each $x\in\ML_\mu$.
\end{lemma}
\begin{proof}
Define the computable martingale
\begin{align*}
d(\epsilon) & = 2\\
d(x_0) & = \frac{1}{\mu[x_0]} \\
d(x_0\ldots x_k) & = \frac{\mu[x_1\ldots x_k]}{\mu[x_0\ldots x_k]}\quad\text{for $k\geq 1$.}
\end{align*}
By the effective Doob's convergence theorem (see Theorem 7.1.3 on page 270 in  \cite{dowhir10}), for each $x\in\ML_\mu$, $d(x_0\ldots x_k)$ converges, and so does $f_k(x)=\log d(x_0\ldots x_k)$.
\end{proof}

Let $f(x)$ be the limit. We write
\begin{equation*}
-\frac{1}{n}\log \mu[x\rst{n}]=\frac{1}{n}\sum_{k=0}^{n-1}(f_{n-1-k}\circ T^k(x)-f\circ T^k(x))+\frac{1}{n}\sum_{k=0}^{n-1}f\circ T^k(x)
\end{equation*}
and prove that the first term tends to $0$ while the second term converges to $\integral{f}{\mu}=h(\mu)$.


We will use the following lemma (Corollary 2.2 on page 261 in \cite{Pet83}, Lemma 4.26 on page 26 in \cite{Smo71}).
\begin{lemma}\label{lem_sup}
$f^*:=\sup_k f_k\in L^1$.
\end{lemma}

As $f_k\to f$ a.e. and the convergence is dominated by $f^*\in L^1$, $f_k\to f$ in $L^1$.

\begin{proposition}
For each $x\in\ML_\mu$,
\begin{equation}\label{eq1}
\lim_n\frac{1}{n}\sum_{k=0}^{n-1}f\circ T^k(x)=\integral{f}{\mu}=h_\mu(P).
\end{equation}
\end{proposition}
\begin{proof}
That $\integral{f}{\mu}=h(\mu)$ is a classical result and follows from $h(\mu)=\lim_k \integral{f_k}{\mu}$ and the $L^1$-convergence of $f_k$ to $f$.

$f^*$ is lower semi-computable and by Lemma \ref{lem_sup} it is a $\mu$-test. By construction, $f$ is $\Delta^0_2$ on $\ML_\mu$ and it is dominated by $f^*$ so it satisfies the conditions of Corollary \ref{cor_delta2_bounded}, from which the result follows directly.
\end{proof}

\begin{proposition}
For each $x\in\ML_\mu$,
\begin{equation}\label{eq2}
\lim_{n\to\infty}\frac{1}{n}\sum_{k=0}^{n-1}f_{n-1-k}\circ T^k(x)-f\circ T^k(x)=0.
\end{equation}
\end{proposition}

\begin{proof}
Let
\begin{equation*}
g_N = \sup_{k\geq N}|f_k-f|\quad\text{and}\quad\tilde{g}_N = \sup_{k,j\geq N}|f_k-f_j|.
\end{equation*}

For $x\in \ML_\mu$,
\begin{align*}
|f_k(x)-f(x)|&=\lim_j |f_k(x)-f_j(x)| \\
& = \limsup_j |f_k(x)-f_j(x)| \\
& \leq \sup_{j\geq N}|f_k(x)-f_j(x)|,
\end{align*}
so ${g}_N(x)\leq \tilde{g}_N(x)$. As $f_k\to f$ a.e., $\tilde{g}_N\to 0$ a.e. As $\tilde{g}_N\leq 2f^*\in L^1$, $\tilde{g}_N\to 0$ in $L^1$ by the dominated convergence theorem.
On $\ML_\mu$,
\begin{align*}
\left|\frac{1}{n}\sum_{k=0}^{n-1}f_{n-1-k}\circ T^k-f\circ T^k\right| & \leq \frac{1}{n}\sum_{k=0}^{n-1}|f_{n-1-k}\circ T^k-f\circ T^k| \\
& = \frac{1}{n}\sum_{k=0}^{n-1-N}|f_{n-1-k}\circ T^k-f\circ T^k|+\frac{1}{n}\sum_{k=n-N}^{n-1}|f_{n-1-k}\circ T^k-f\circ T^k|\\
& \leq \frac{1}{n}\sum_{k=0}^{n-1-N}g_N\circ T^k+\frac{1}{n}\sum_{k=n-N}^{n-1}(f^*+f)\circ T^k \\
& \leq \frac{1}{n}\sum_{k=0}^{n-1-N}\tilde{g}_N\circ T^k+\frac{1}{n}\sum_{k=0}^{n-1}(f^*+f)\circ T^k-\frac{1}{n}\sum_{k=0}^{n-N-1}(f^*+f)\circ T^k.
\end{align*}
Fix $N$ and let $n\to\infty$. As $\tilde{g}_N\in L^1$ is lower semi-computable, the first term converges to $\integral{\tilde{g}_N}{\mu}$ by the Effective Ergodic Theorem \ref{thm_erg_semi}. As $f^*+f$ is $\Delta^0_2$ on $\ML_\mu$ and is dominated by the $\mu$-test $2f^*$, the second and the third terms converge to $\integral{(f^*+f)}{\mu}$ by Corollary \ref{cor_delta2_bounded} so their limits cancel each other.

As $\integral{\tilde{g}_N}{\mu}\to 0$, we have proved equality (\ref{eq2}).
\end{proof}

Putting equalities (\ref{eq0}), (\ref{eq1}) and (\ref{eq2}) together gives, for $x\in\ML_\mu$,
\begin{equation*}
\lim_n-\frac{1}{n}\log \mu[x\rst{n}] = \lim_n\frac{1}{n}\sum_{k=0}^{n-1}f_{n-1-k}\circ T^{k}(x)=h(\mu).
\end{equation*}

\newcommand{\etalchar}[1]{$^{#1}$}


\begin{thebibliography}{FGMN10}

\bibitem[AHLM07]{Ath07}
Krishna~B. Athreya, John~M. Hitchcock, Jack~H. Lutz, and Elvira Mayordomo.
\newblock Effective strong dimension in algorithmic information and
  computational complexity.
\newblock {\em SIAM J. Comput.}, 37(3):671--705, 2007.

\bibitem[BDH{\etalchar{+}}10]{BieHoyShen10}
Laurent Bienvenu, Adam~R. Day, Mathieu Hoyrup, Ilya Mezhirov, and Alexander
  Shen.
\newblock A constructive version of {B}irkhoff's ergodic theorem for
  {M}artin-{L}{\"o}f random points.
\newblock Submitted. ArXiv 1007.5249, 2010.

\bibitem[BDMS10]{BienvenuDMS10}
Laurent Bienvenu, Adam Day, Ilya Mezhirov, and Alexander Shen.
\newblock Ergodic-type characterizations of algorithmic randomness.
\newblock In {\em Computability in Europe (CIE 2010)}, volume 6158 of {\em
  Lecture Notes in Computer Science}, pages 49--58. Springer, 2010.

\bibitem[Cha75]{Cha75}
Gregory~J. Chaitin.
\newblock A theory of program size formally identical to information theory.
\newblock {\em J. ACM}, 22(3):329--340, 1975.

\bibitem[DH10]{dowhir10}
Rod Downey and Denis Hirschfeldt.
\newblock {\em Algorithmic Randomness and Complexity}.
\newblock Springer-Verlag New York, Inc., Secaucus, NJ, USA, 2010.

\bibitem[FGMN10]{FraGreeMillNg10}
Johanna~N.Y. Franklin, Noam Greenberg, Joseph~S. Miller, and {Keng Meng} Ng.
\newblock {M}artin-{L}{\"o}f random points satisfy {B}irkhoff's ergodic theorem
  for effectively closed sets.
\newblock To appear in the \emph{Proceedings of the American Mathematical
  Society}, 2010.

\bibitem[G{\'a}c80]{Gac80}
P{\'e}ter G{\'a}cs.
\newblock Exact expressions for some randomness tests.
\newblock {\em Z. Math. Log. Grdl. M.}, 26:385--394, 1980.

\bibitem[Lev73]{Lev73}
Leonid~A. Levin.
\newblock On the notion of a random sequence.
\newblock {\em Soviet Mathematics Doklady}, 14:1413--1416, 1973.

\bibitem[Lut00]{Lutz00}
Jack~H. Lutz.
\newblock Dimension in complexity classes.
\newblock In {\em IEEE Conference on Computational Complexity}, pages 158--169,
  2000.

\bibitem[Lut05]{Lutz05}
Jack~H. Lutz.
\newblock Effective fractal dimensions.
\newblock {\em Mathematical Logic Quarterly}, 51(1):62--72, 2005.

\bibitem[May02]{May02}
Elvira Mayordomo.
\newblock A {K}olmogorov complexity characterization of constructive
  {H}ausdorff dimension.
\newblock {\em Inf. Process. Lett.}, 84(1):1--3, 2002.

\bibitem[ML66]{MLof66}
Per Martin-L{\"o}f.
\newblock The definition of random sequences.
\newblock {\em Information and Control}, 9(6):602--619, 1966.

\bibitem[Nak05]{Nak05}
Masahiro Nakamura.
\newblock Ergodic theorems for algorithmically random sequences.
\newblock {\em Proceedings of the Symposium on Information Theory and Its
  Applications}, 2005.

\bibitem[Nie09]{Nies09}
A.~Nies.
\newblock {\em Computability and randomness}.
\newblock Oxford logic guides. Oxford University Press, 2009.

\bibitem[Pet83]{Pet83}
Karl Petersen.
\newblock {\em Ergodic Theory}.
\newblock Cambridge Univ. Press, 1983.

\bibitem[Sch73]{Sch73}
Claus-Peter Schnorr.
\newblock Process complexity and effective random tests.
\newblock {\em J. Comput. Syst. Sci.}, 7(4):376--388, 1973.

\bibitem[Smo71]{Smo71}
Meir Smorodinsky.
\newblock {\em Ergodic Theory, Entropy}, volume 214 of {\em Lecture Notes in
  Mathematics}.
\newblock Springer-Verlag, Berlin Heibelberg New York, 1971.

\bibitem[V'y98]{Vyu98}
Vladimir~V. V'yugin.
\newblock Ergodic theorems for individual random sequences.
\newblock {\em Theoretical Computer Science}, 207(4):343--361, 1998.

\end{thebibliography}
\end{document}